\documentclass[10pt,twocolumn]{article}
\usepackage[margin=0.75in]{geometry}
\usepackage{amsmath,amssymb}
\usepackage{graphicx}
\usepackage{url}
\usepackage[hyphens]{xurl}
\usepackage[colorlinks=true,linkcolor=blue,citecolor=blue,urlcolor=blue]{hyperref}
\usepackage{booktabs}
\usepackage{titlesec}
\usepackage{tikz}
\usepackage{microtype}
\usetikzlibrary{arrows.meta,positioning}

\titleformat{\section}{\normalfont\large\bfseries}{\thesection}{1em}{}
\titleformat{\subsection}{\normalfont\normalsize\bfseries}{\thesubsection}{1em}{}

\title{\textbf{\Large A Browser-Based Gesture-Driven Avatar Interaction Framework for Metaverse Onboarding Environments}}
\author{
Deepti Parachuri$^1$ \quad Chhayank Sahu$^1$ \quad Sameer Singh Choudhary$^1$ \\
\normalsize $^1$Infosys Limited \\
\normalsize \texttt{\{deepti\_parachuri, chhayank.sahu, sameer\_choudhary\}@infosys.com}
}
\date{}

\begin{document}
\maketitle

\begin{abstract}
\noindent Avatar interaction shapes how engaging and immersive a metaverse experience feels, and for that interaction to feel natural, avatars need to respond to users without forcing them through a controller-based interface first. This paper describes a gesture-driven interaction layer built for a browser-based metaverse onboarding environment, where users explore a set of virtual rooms as an avatar and interact with embedded video, document, and quiz content using hand, arm, and head gestures instead of a keyboard or controller. The system combines real-time gesture recognition (Google MediaPipe) with two alternative locomotion techniques---hand-raise navigation and in-place walking---so users can trade off precision against physical immersion depending on the task. The contribution is the integration, deployment, and evaluation of these techniques as a single lightweight, web-deployable, controller-free interaction model, assessed through a structured internal onboarding session with five participants. We report what worked, what didn't, and the design trade-offs that came out of combining these techniques in one deployed system.

\vspace{0.5em}
\end{abstract}
\vspace{1em}

\section{Introduction}

The metaverse is getting attention as a shared digital space where people interact, collaborate, and move through virtual environments using avatars. For that to feel natural, the interaction method matters as much as the visuals. Most current systems still rely on keyboards, mice, or handheld controllers. These are familiar, but they break immersion---users are managing a device instead of just acting.

Gestures are how people already communicate. Hand movement, posture, and body language carry intent without anyone having to think about it. Bringing that into a virtual environment, instead of asking users to learn a control scheme first, is a reasonable way to lower the barrier to entry for first-time users.

Hand-raise navigation, walking-in-place, and dwell-based selection have all been studied in the VR literature for over a decade \cite{bowman2001, steinicke2013, nilsson2013tapping}, and MediaPipe itself is an existing, published hand-tracking framework \cite{zhang2020mediapipe}. The novelty of this paper lies in the integration, deployment, and evaluation of these techniques as a single lightweight, browser-based, controller-free interaction framework for metaverse onboarding: combining vision-based gesture recognition with two alternative locomotion strategies inside a real deployed environment, without requiring dedicated sensors, wearables, or a native VR client, and reporting what we learned deploying it with real users.

The contributions are:

\begin{itemize}
\item A gesture-driven interaction framework for controller-free avatar navigation and content interaction in a browser-based metaverse environment.
\item Two alternative locomotion techniques---hand-based navigation and in-place walking---implemented side by side so users can choose based on task and comfort.
\item An integration of real-time gesture recognition (Google MediaPipe) for navigation, pointer control, and content interaction, deployed without specialized hardware.
\item An account of a structured internal evaluation with five participants during a scheduled onboarding session, reporting where the approach worked and where it didn't.
\end{itemize}

\section{Related Work}

Early virtual environments relied on keyboards, mice, and game controllers for avatar navigation \cite{bowman2001}. These work, but they put a layer of device-mapping between what the user intends and what the avatar does, which costs immersion.

VR interaction moved toward motion-tracked controllers, letting users perform spatial actions through direct hand movement \cite{jerald2015}. This improved responsiveness but still required learning button mappings and kept users tied to specific hardware---interaction was shaped by what the device could do, not by natural human movement.

Gesture-based interaction was explored as an alternative. Prior work used hand gestures, arm movement, and full-body motion to control avatars and manipulate virtual objects \cite{bowman2005, kasahara2013extouch}, and found that mapping real-world movement to virtual action improves engagement and strengthens embodiment. Much of this work, though, depended on specialized sensors, wearables, or depth cameras, which limits how easily it can be deployed at scale.

Expressive avatars matter beyond navigation. Gesture-driven non-verbal behavior, posture, and body language improve social presence and realism in avatar-mediated interaction \cite{pan2016}, which is directly relevant to collaborative and learning-oriented metaverse spaces---though expressive interaction usually comes with more system complexity and calibration overhead.

Locomotion is its own open problem. There is a persistent trade-off between navigation precision, immersion, and physical comfort \cite{steinicke2013}: joystick-style navigation is precise but not very embodied, while natural walking is embodied but tiring and constrained by physical space. Walking-in-place has been proposed as a middle ground that keeps a sensation of walking while the user stays stationary \cite{nilsson2013tapping, nilsson2018review}.

More recently, web-based XR and vision-based tracking have made controller-free interaction possible without dedicated hardware. MediaPipe \cite{zhang2020mediapipe} enables real-time hand and body tracking from camera input alone, which is what makes browser-based gesture interaction practical in the first place. Integrating this kind of tracking into a large, interactive metaverse space while keeping interaction stable and responsive is still an open problem \cite{speicher2019}.

What we describe here differs from this prior work mainly in scope and deployment target, not in technique: a lightweight, browser-based system that combines vision-based gesture recognition with two locomotion strategies inside one deployed onboarding environment, prioritizing accessibility and ease of deployment over sensing fidelity.

\section{System Overview}

The system is a browser-based metaverse environment used for onboarding: users enter as a chosen avatar and explore a set of themed virtual rooms containing embedded video, document, and quiz content.

On entry, users are represented by a personalized avatar. An avatar buddy accompanies new users, gives a short overview of the space, and explains the available interaction methods, which reduces the amount of up-front instruction needed before someone can start exploring on their own.

The environment is organized into rooms, each built around a topic or use case. Within a room, users interact with virtual objects: television screens playing video, panels showing PDF documents, and interactive quiz elements. This spatial, room-based structure ties content to a location rather than a menu.

Interaction is gesture-driven throughout. Users navigate, select content, and interact with objects using hand movement, arm motion, and head orientation, tracked in real time via Google MediaPipe. Recognized gestures map to navigation, rotation, selection, and content-control actions.

The system runs on web-based XR technology, accessible through a standard browser without a dedicated VR headset or client install. The architecture is modular, so new rooms, content types, and gestures can be added without touching the core pipeline.

Although built for this specific onboarding deployment, the interaction framework itself is not domain-specific. The same gesture design could apply to virtual training, collaborative workspaces, or other browser-based immersive applications.

\section{System Architecture and Data Flow}

\begin{figure}[t]
\centering
\begin{tikzpicture}[
    node distance=0.45cm and 0cm,
    box/.style={rectangle, draw, rounded corners, minimum width=4.6cm, minimum height=0.7cm, align=center, font=\small},
    arr/.style={-{Latex[length=2mm]}, thick}
]
\node[box] (cam) {Input Layer\\\footnotesize RGB Camera};
\node[box, below=of cam] (mp) {Perception Layer\\\footnotesize Vision-Based Tracking (MediaPipe)\\\footnotesize Hand, Body, Head Landmarks};
\node[box, below=of mp] (interp) {Interpretation Layer\\\footnotesize Gesture Detection \& Locomotion Intent};
\node[box, below=of interp] (ctrl) {Control Layer\\\footnotesize Interaction Controller\\\footnotesize Navigation, Pointer, Content};
\node[box, below=of ctrl] (meta) {Metaverse Layer\\\footnotesize Avatar / Camera Controller\\\footnotesize Virtual Rooms \& Media Content};

\draw[arr] (cam) -- (mp);
\draw[arr] (mp) -- (interp);
\draw[arr] (interp) -- (ctrl);
\draw[arr] (ctrl) -- (meta);
\end{tikzpicture}
\caption{Layered system architecture. Camera input is processed into landmarks, interpreted as gestures and locomotion intent, mapped to interaction actions, and applied to the avatar and virtual environment.}
\label{fig:architecture}
\end{figure}
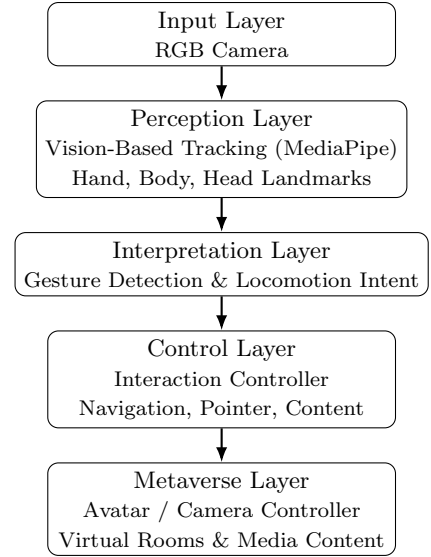

Figure~\ref{fig:architecture} shows the architecture as a layered pipeline that separates gesture capture, interpretation, and interaction control. A standard RGB camera feeds a perception layer (vision-based tracking via Google MediaPipe), which produces hand, body, and head landmarks. These are passed to an interpretation layer that detects gestures and resolves locomotion intent, then to a control layer (an interaction controller for navigation, pointer control, and content interaction), which drives the metaverse layer (avatar and camera control, virtual rooms, and media content).

A standard RGB camera captures user input, processed through a vision-based tracking layer. Hand, body, and head landmarks are extracted and passed to a gesture interpretation layer, which detects navigation gestures, locomotion intent, and pointer-based interaction. An interaction controller maps recognized gestures to avatar actions---movement, camera rotation, content selection---applied to the scene in real time. Decoupling recognition from interaction logic is what lets the system support multiple locomotion techniques and interaction modes without a rewrite each time.

\subsection{End-to-End Interaction Workflow}
A typical example: a user raises their right hand while facing a target room. The camera captures the motion, hand landmarks are extracted, and the interpretation layer detects a forward-navigation gesture, resolving direction from the current head orientation. The interaction controller then drives continuous avatar movement until the gesture is released. If the user switches to a pointing gesture instead, navigation mode deactivates and pointer interaction takes over, allowing content selection. Perception, interpretation, and control operate as a single pipeline, so this transition happens without an explicit mode switch from the user.

\section{Gesture-Driven Interaction Design}

The gesture set is split into two categories---movement and pointer interaction---so users have a clear line between navigating and interacting with content, which keeps cognitive load down.

\subsection{Movement Gestures}
Table~\ref{tab:movement} lists the movement gestures. Raising the right hand (fingers extended) moves the avatar forward; raising the left hand moves it backward. This symmetric mapping is easy to remember. Arm swinging is recognized as an alternative forward-movement trigger, mimicking a walking motion for a more embodied, hands-free option. Head orientation controls rotation directly: turning left or right rotates the view, without a separate rotation gesture.

\begin{table}[t]
\centering
\small
\caption{Movement gestures and avatar navigation actions.}
\label{tab:movement}
\begin{tabular}{p{2.1cm}p{2.1cm}p{3cm}}
\toprule
\textbf{Gesture} & \textbf{Action} & \textbf{Description} \\
\midrule
Open right hand & Move forward & Raise right hand, fingers extended \\
Open left hand & Move backward & Raise left hand, fingers extended \\
Arm swinging & Walk forward & Swing both arms to simulate walking \\
Head turn left & Rotate view right & Turn head toward the left \\
Head turn right & Rotate view left & Turn head toward the right \\
\bottomrule
\end{tabular}
\end{table}

\subsection{Locomotion Techniques}
Two locomotion techniques were implemented to balance ease of use against immersion.

\textbf{Hand-and-head navigation.} Raise the right hand to move forward, the left to move backward; head rotation sets direction. Low physical effort, precise, well suited to short exploration and content-focused tasks.

\textbf{In-place walking.} Users perform a walking motion while stationary; rhythmic body movement triggers forward motion, and head rotation sets direction. This is meant to create a stronger sense of embodiment at the cost of more physical effort.

Supporting both lets the system adapt to what a user actually wants at that moment---quick, precise access to content, or a more physically engaged walkthrough.

\subsection{Pointer-Based Interaction}
Table~\ref{tab:pointer} lists the pointer gestures, used for interacting with UI elements and content objects.

\begin{table}[t]
\centering
\small
\caption{Pointer-based interaction gestures.}
\label{tab:pointer}
\begin{tabular}{p{2.1cm}p{2.1cm}p{3cm}}
\toprule
\textbf{Gesture} & \textbf{Action} & \textbf{Description} \\
\midrule
Pointing & Activate pointer & Extend index finger, other fingers closed \\
Hover 3s & Select element & Hold pointer over a UI element for 3 seconds \\
Swipe up/down & Scroll content & Move pointing hand vertically \\
\bottomrule
\end{tabular}
\end{table}

Extending the index finger while keeping other fingers closed activates a virtual pointer, used to target video screens, quiz panels, and document viewers.

\subsection{Design Rationale}
The gesture set favors simple, easily distinguishable gestures over complex multi-finger combinations, since that improves recognition robustness and reduces fatigue. Supporting two locomotion techniques is a direct response to the precision-versus-immersion trade-off discussed in Section II, rather than an attempt to find one universally best technique.

\section{Implementation}

The system is implemented with web-based XR technology for browser and device independence, with gesture recognition and avatar control running in real time.

\subsection{Gesture Recognition Pipeline}
Google MediaPipe \cite{zhang2020mediapipe} tracks hand, body, and head landmarks from camera input---finger joints, palm orientation, shoulder position, head direction. Hand landmark configurations identify raised-hand and pointing gestures; temporal changes in body landmarks identify walking motion; head landmarks drive camera rotation and movement direction.

\subsection{Locomotion Implementation}
For hand-based locomotion, finger extension and palm height relative to the upper body determine whether a raised-hand gesture is active, with the right hand triggering forward motion and the left triggering backward. Head orientation is tracked continuously to set direction.

For in-place walking, body landmark motion is analyzed over time to detect rhythmic patterns consistent with walking; once detected, forward movement is triggered, with head rotation again setting direction. A modular control layer switches between the two locomotion methods, which also makes it straightforward to add further navigation techniques later.

\subsection{Interaction Stability and Gesture Control}
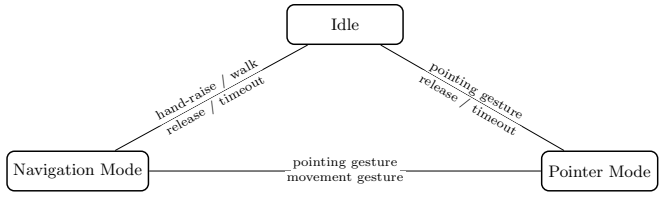
\begin{figure}[t]
\centering
\resizebox{0.99\linewidth}{!}{%
\begin{tikzpicture}[
  node distance=2.0cm and 2.6cm,
  state/.style={rectangle, rounded corners, draw=black, thick, minimum width=2.2cm, minimum height=0.75cm, align=center, font=\small},
  every edge/.style={draw, {Latex[length=2mm]}-{Latex[length=2mm]}, thick},
  lbl/.style={font=\scriptsize, align=center, fill=white, inner sep=1pt}
]
\node[state] (idle) {Idle};
\node[state, below left=of idle] (nav) {Navigation Mode};
\node[state, below right=of idle] (ptr) {Pointer Mode};

\draw (idle) to node[lbl, sloped, above] {hand-raise / walk} node[lbl, sloped, below] {release / timeout} (nav);
\draw (idle) to node[lbl, sloped, above] {pointing gesture} node[lbl, sloped, below] {release / timeout} (ptr);
\draw (nav) to node[lbl, above] {pointing gesture} node[lbl, below] {movement gesture} (ptr);
\end{tikzpicture}%
}
\caption{Gesture-interaction state machine. Each edge is bidirectional: the upper label triggers the transition away from Idle (or between Navigation and Pointer modes), and the lower label triggers the return. Navigation and pointer modes are mutually exclusive; a cooldown after each transition (not shown) suppresses repeated triggering.}
\label{fig:statemachine}
\end{figure}

Gesture recognition uses temporal consistency across multiple frames rather than single-frame detection, with thresholds on landmark position and movement to reduce false positives during transitional gestures. A short cooldown after gesture activation prevents repeated or unintended commands. Navigation and pointer interaction are kept as distinct states, with only one active at a time, to avoid ambiguity between locomotion and content interaction. Figure~\ref{fig:statemachine} summarizes this as a state diagram: the system idles until a movement or pointing gesture is detected, and navigation and pointer interaction remain mutually exclusive until an explicit mode-switching gesture or a timeout returns the system to idle.

\begin{figure}[t]
\centering
\includegraphics[width=0.95\linewidth]{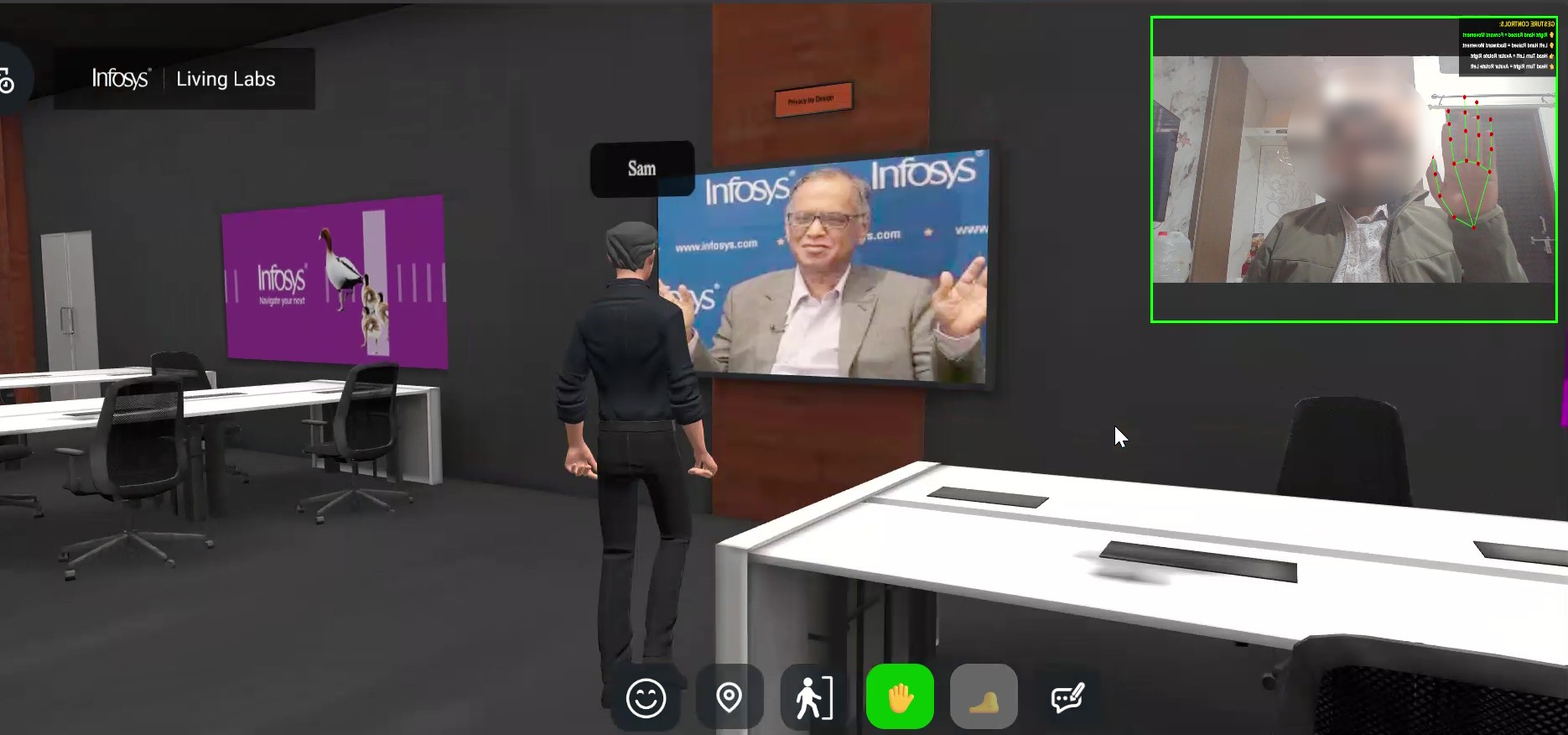}
\caption{Hand and head based locomotion in the onboarding environment. Head rotation controls camera direction, while raised-hand gestures drive forward and backward movement. The participant's face in the tracking inset is blurred for privacy; the hand-landmark overlay used for gesture recognition remains visible.}
\label{fig:handhead}
\end{figure}

\begin{figure}[t]
\centering
\includegraphics[width=0.95\linewidth]{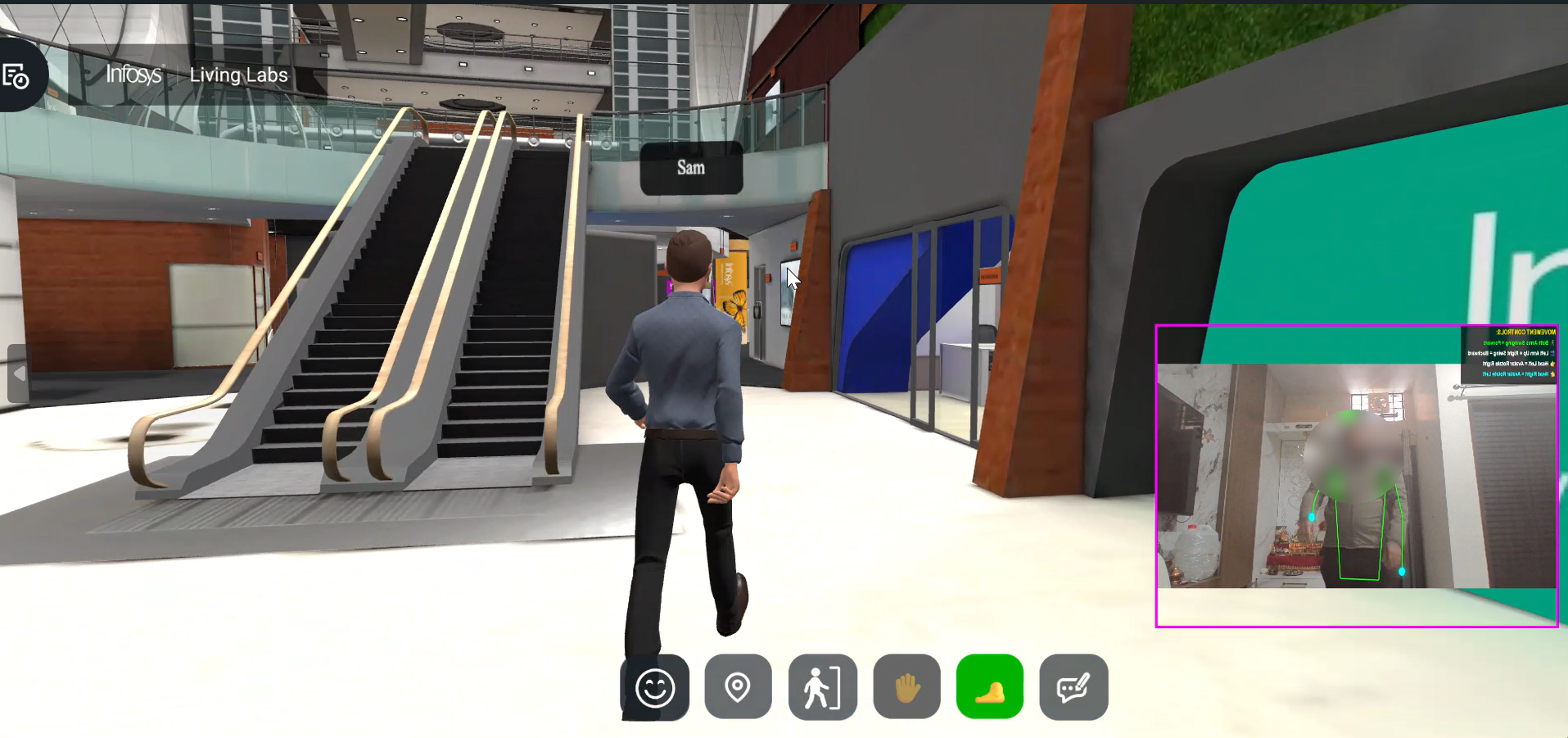}
\caption{In-place walking locomotion using body landmark tracking. Rhythmic body movement initiates forward motion, while head orientation controls navigation direction. The participant's face is blurred for privacy; the body-landmark overlay used for gesture recognition remains visible.}
\label{fig:inplacewalk}
\end{figure}

\section{Evaluation}

\subsection{Method}
We evaluated the system with five participants ($n=5$) during a structured internal onboarding event. Each session ran approximately 30 minutes and followed the same sequence: avatar selection and orientation with the avatar buddy, navigation between two or more rooms, interaction with embedded content (video, document, quiz), and open use of both locomotion techniques. This was an internal, qualitative evaluation rather than a controlled study---we did not run a between-technique statistical comparison, and we report it as such rather than overstating what the sample size and format support.

\subsection{Observations}
Participants entered the environment by selecting an avatar and were oriented by the avatar buddy. This guidance was reported as helpful for reducing initial confusion, letting participants start exploring with little additional assistance.

Movement gestures were picked up quickly, particularly open-hand movement and arm swinging, which participants described as aligning naturally with real-world action. Head-based rotation was also perceived as intuitive, letting people look around without a separate control.

Pointer-based interaction---pointing combined with dwell-based selection---supported engagement with videos, documents, and quiz elements without a physical controller. Participants noted that the dwell-based mechanism reduced accidental selection and gave a clear sense of control. Swipe gestures for scrolling were picked up without difficulty.

The room-based layout helped participants keep track of where content was and what it covered; interacting with a screen or document in its physical location felt more contextual than navigating a menu.

Two issues came up consistently. First, gesture recognition sometimes needed more deliberate movement to register reliably, especially under inconsistent lighting. Second, prolonged use of in-place walking led to mild fatigue for some participants, which is consistent with what the locomotion literature already reports about walking-based techniques \cite{steinicke2013}.

Participants distinguished clearly between the two locomotion techniques: hand-based navigation was easy to learn and low-effort, better suited to short, content-focused tasks; in-place walking felt more immersive and physically connected, but tired people out faster. Neither technique was preferred outright---people wanted both, for different moments in the session.

Representative comments included: ``I didn't have to figure out controls, the gestures just worked''; ``the walking felt more real, but my arms got tired after a while''; and ``pointing at the screen to select things felt natural.''

\section{Design Principles and Lessons Learned}

A few practical points came out of building and running this system.

Simple, clearly distinguishable gestures worked better than complex combinations, both for recognition robustness and for how quickly people picked them up. Nobody needed more than a brief explanation to use the core gesture set.

No single locomotion technique was best for everything. Hand-based navigation is precise and low-effort; in-place walking is more immersive but more tiring. Offering both, rather than picking one, let the system adapt to what a given task actually needed.

Keeping navigation and content-interaction states clearly separate mattered more than we expected going in. When the two modes could overlap, users hesitated and made more accidental inputs; keeping them mutually exclusive made behavior more predictable and users more confident.

\subsection{Design Trade-offs}
More expressive gestures generally hurt recognition robustness and learnability, so the system leans toward a small, simple set rather than trying to cover a wide gesture vocabulary. Similarly, immersion and physical comfort pull in opposite directions for locomotion---which is why the system supports two techniques instead of committing to one.

\section{Limitations and Future Work}

The evaluation reported here is small ($n=5$), internal, and qualitative; it tells us the approach is usable and where friction shows up, not how it performs against a controller-based baseline or at statistically meaningful scale. A larger, controlled comparison---ideally against a menu-driven baseline, in the style of prior HAI evaluation work---is the natural next step.

We also did not collect quantitative system-performance data (gesture-recognition latency, tracking frame rate, task completion time, or comparative error rates against keyboard/controller input) during this deployment. These are standard measures for a systems paper in this space, and their absence limits how precisely the runtime behavior described in Sections IV--VI can be characterized. Reporting these figures under controlled, repeatable conditions is planned future work rather than something we estimate here.

The system depends on camera-based gesture recognition, which is sensitive to lighting and camera placement, and occasionally needed more deliberate gestures for reliable detection. In-place walking, in particular, can cause physical fatigue over longer sessions.

Future work includes improving recognition robustness, adding multimodal input such as voice commands, supporting multi-user collaborative sessions, and running a larger, formal user study with quantitative usability and immersion measures.

\section{Conclusion}

We presented a gesture-driven avatar interaction framework for a browser-based metaverse onboarding environment, combining real-time hand, body, and head gesture recognition with two locomotion techniques and pointer-based content interaction. The contribution is the integration, deployment, and evaluation of these established techniques as a single lightweight, web-deployable interaction model, tested with real users in a structured onboarding session. A small internal evaluation ($n=5$) suggests the approach is usable and reduces reliance on traditional input devices, with a clear trade-off between the precision of hand-based navigation and the immersion of in-place walking. The framework is not tied to this specific onboarding use case, and the same interaction design could extend to training, collaboration, or other browser-based immersive applications. We see a larger, controlled evaluation as the most useful next step for establishing how this compares to controller-based interaction more rigorously.


\begin{thebibliography}{99}
\small
\setlength{\itemsep}{0pt}
\setlength{\parskip}{0pt}

\bibitem{bowman2001} D. A. Bowman, L. F. Hodges, J. L. Bolter, and A. MacIntyre, ``Formalizing the design, evaluation, and application of interaction techniques for immersive virtual environments,'' \emph{Journal of Visual Languages and Computing}, vol. 12, no. 4, pp. 335--355, Aug. 2001.

\bibitem{jerald2015} J. Jerald, \emph{The VR Book: Human-Centered Design for Virtual Reality}. San Rafael, CA: Morgan \& Claypool, 2015.

\bibitem{slater1997} M. Slater and S. Wilbur, ``A framework for immersive virtual environments (FIVE): Speculations on the role of presence in virtual environments,'' \emph{Presence: Teleoperators and Virtual Environments}, vol. 6, no. 6, pp. 603--616, Dec. 1997.

\bibitem{pan2016} Y. Pan and A. Steed, ``A comparison of avatar-, video-, and robot-mediated interaction on users' sense of embodiment,'' \emph{IEEE Transactions on Visualization and Computer Graphics}, vol. 22, no. 4, pp. 1250--1259, Apr. 2016.

\bibitem{bowman2005} D. A. Bowman, E. Kruijff, J. J. LaViola, and I. Poupyrev, \emph{3D User Interfaces: Theory and Practice}. Boston, MA: Addison-Wesley, 2005.

\bibitem{steinicke2013} F. Steinicke, Y. Visell, J. Campos, and A. L\'ecuyer, Eds., \emph{Human Walking in Virtual Environments: Perception, Technology, and Applications}. New York, NY: Springer, 2013.

\bibitem{kasahara2013extouch} S. Kasahara, R. Niiyama, V. Heun, and H. Ishii, ``exTouch: Spatially-aware embodied manipulation of actuated objects mediated by augmented reality,'' in \emph{Proc. 7th Int. Conf. Tangible, Embedded and Embodied Interaction (TEI)}, 2013, pp. 223--228.

\bibitem{lugrin2015} J.-L. Lugrin, D. Landeck, and M. E. Latoschik, ``Avatar embodiment realism and virtual body ownership,'' in \emph{Proc. IEEE Virtual Reality (VR)}, 2015.

\bibitem{zhang2020mediapipe} F. Zhang, V. Bazarevsky, A. Vakunov, A. Tkachenka, G. Sung, C.-L. Chang, and M. Grundmann, ``MediaPipe Hands: On-device real-time hand tracking,'' \emph{arXiv preprint arXiv:2006.10214}, 2020.

\bibitem{nilsson2018review} N. C. Nilsson, S. Serafin, F. Steinicke, and R. Nordahl, ``Natural walking in virtual reality: A review,'' \emph{Computers in Entertainment}, vol. 16, no. 2, Art. 8, 2018.

\bibitem{nilsson2013tapping} N. C. Nilsson, S. Serafin, M. H. Laursen, K. S. Pedersen, E. Sikstr\"om, and R. Nordahl, ``Tapping-in-place: Increasing the naturalness of immersive walking-in-place locomotion through novel gestural input,'' in \emph{Proc. IEEE Symposium on 3D User Interfaces (3DUI)}, 2013, pp. 31--38.

\bibitem{speicher2019} M. Speicher, B. D. Hall, and M. Nebeling, ``What is mixed reality?'' in \emph{Proc. CHI Conference on Human Factors in Computing Systems}, 2019.

\end{thebibliography}
\end{document}